\documentclass[
twocolumn, preprintnumbers, amsmath,amssymb,nofootinbib]{revtex4}

\usepackage{graphicx}
\usepackage{dcolumn}
\usepackage{bm}

\newcommand\ee{\end{equation}}
\newcommand\be{\begin{equation}}
\newcommand\eea{\end{eqnarray}}
\newcommand\bea{\begin{eqnarray}}

\newcommand\lsim{\mathrel{\rlap{\lower4pt\hbox{\hskip1pt$\sim$}}
        \raise1pt\hbox{$<$}}}
\newcommand\gsim{\mathrel{\rlap{\lower4pt\hbox{\hskip1pt$\sim$}}
        \raise1pt\hbox{$>$}}}

\begin{document}


\title{An observational test of the Vainshtein mechanism
}

\author{Lam Hui}
\email{lhui@astro.columbia.edu}

\author{Alberto Nicolis}
\email{nicolis@phys.columbia.edu}

\affiliation{%
Physics Department and Institute for Strings, Cosmology, and Astroparticle Physics,\\
Columbia University, New York, NY 10027, USA
}%

\date{\today}

\begin{abstract}
Modified gravity theories capable of genuine self-acceleration typically invoke a
galileon scalar which mediates a long range force, but is
screened by the Vainshtein mechanism on small scales. In such theories, 
non-relativistic stars carry the full scalar charge (proportional to their 
mass), while black holes carry none. Thus, for a galaxy free-falling in some
external gravitational field, its central massive black hole is expected to lag behind
the stars. To look for this effect, and to distinguish it from other
astrophysical effects, one can correlate the
gravitational pull from the surrounding structure
with the offset between the stellar center and the black hole.
The expected offset depends on the central density of the galaxy,
and ranges up to $\sim 0.1$ kpc for small galaxies. The observed offset in M87
cannot be explained by this effect unless the scalar force is
significantly stronger than gravity. We also discuss the systematic
offset of compact objects from the galactic plane as another
possible signature.
\end{abstract}

\maketitle


There has been a lot of interest in theories of modified gravity that
might explain the observed accelerated expansion of the universe \cite{Bhuvreview,Kurtreview}.
Theories capable of genuine self-acceleration --
without invoking vacuum energy in
the Einstein frame \cite{HKW} -- are especially interesting, and they generally
involve introducing a scalar ($\varphi$) that respects the galileon
symmetry $\varphi \rightarrow \varphi + b + c_\mu x^\mu$, where
$b$ is a constant and $c_\mu$ is a constant vector \cite{NRT}.
This scalar mediates a long-ranged force, the
so-called fifth force in addition to the usual gravitational force
between objects. Thanks to the Vainshtein mechanism \cite{Vainshtein}, the scalar is screened on small scales, so that solar system constraints
are satisfied. To see how it works, let us illustrate with the
simplest galileon model, inspired by DGP \cite{DGP,LPR}; the
equation
for the scalar (in Einstein frame) is
\footnote{Here, $\varphi$ is related to the standard notation $\pi$ for the galileon
by $\pi = \alpha \varphi $.}
\footnote{Two additional galileon symmetric interaction
terms can be written down in the equation of motion:
$(\Box \varphi)^3 - 3 \Box \varphi (\partial_\mu\partial_\nu\varphi)^2
+ 2 (\partial_\mu\partial_\nu\varphi)^3$ and
$(\Box \varphi)^4 - 6 (\Box \varphi)^2
(\partial_\mu\partial_\nu\varphi)^2
+ 8 \Box \varphi (\partial_\mu\partial_\nu\varphi)^3 +
3 [(\partial_\mu\partial_\nu\varphi)^2]^2 - 6
(\partial_\mu\partial_\nu\varphi)^4$. 
All results in this paper apply in the presence of
any combination of the
galileon terms.}:
\begin{eqnarray}
\label{eom}
\Box \varphi +  {2\over 3 m^2} \left[ (\Box \varphi)^2
  - \partial_\mu \partial_\nu \varphi \partial^\mu \partial^\nu
  \varphi \right] =  - 8\pi \alpha G \, {T}_\mu {}^\mu \, ,
\end{eqnarray}
where $T_\mu {}^\mu$ is the trace of the matter
energy-momentum, which for our purpose
can be equated with (negative of)
the matter density $\rho$, assuming it is non-relativistic.
The constant $\alpha$ quantifies the scalar-matter coupling,
and is generically of order unity, i.e.~of gravitational strength.
For instance, it takes
the value $1/\sqrt{6}$ in massive gravity models
\cite{CGT}. 
The mass scale $m$ is generally of the order
of the Hubble constant today $m \sim H_0$.
We are interested in solutions of this equation in
the quasi-static limit, meaning time derivatives can be ignored
and $\Box \rightarrow \nabla^2$.
On large scales, the linear term $\nabla^2 \varphi$ dominates;
Eq. (\ref{eom}) resembles the Poisson equation, with
$\varphi$ playing the role of the gravitational potential.
A localized source $\rho$ yields a profile $\varphi$ that
scales inversely with distance $r$.
However, as one approaches the source,
the interaction term (second term on the l.h.s.) dominates,
and simple power counting reveals $\varphi \propto \sqrt{r}$.
Thus, at small distances, $\varphi$ is screened relative
to the normal $1/r$ gravitational potential.
The transition scale is known as the Vainshtein radius,
and is roughly given by $(GM/m^2)^{1/3}$ for spherically symmetric
configurations, where $M$ is the mass of the source.
For instance, the entire solar system fits within
the Vainshtein radius of the Sun, about $0.1$ kpc, greatly
suppressing the scalar force sourced by the Sun.
What makes the galileon model attractive from this viewpoint, is that it is
the same nonlinear interaction that is responsible both for
Vainshtein screening, and for self-acceleration \cite{NRT}.

An important property of Eq. (\ref{eom}) is that
it can be rewritten in the form $\partial_\mu J^\mu = - T_\mu
{}^\mu$, where $J^\mu$ is a nonlinear function of derivatives of
$\varphi$ \cite{NRT}. One can thus define a scalar ``charge'' 
$Q = - \int d^3 x \, T_\mu {}^\mu = \int d^3 x \,\rho$,
\footnote{One might wonder 
if $-\int d^3 x T_\mu {}^\mu = \int d^3 x \, \rho$ is too strong an
assumption, since an apparently non-relativistic object might
have relativistic components, e.g. gluons in protons.
It turns out that as long as the object in question is stationary,
and has a finite extent, $-\int d^3 x T_\mu {}^\mu = - \int d^3 x T_0
{}^0 = \int d^3 x \, \rho$ holds, by virtue of a tensorial virial
theorem \cite{us2}. 
For further discussions, including 
the renormalization of $Q$ by quantum effects, see
\cite{us2,cristian}. 
}
which is none other than the mass $M$.
\footnote{That is, aside from an exception described below.}
As shown in \cite{us}, the scalar charge
also quantifies the response of an object to an external
field, i.e. an external gravitational $\Phi_{\rm ext}$ + scalar
$\varphi_{\rm ext}$ field exerts a net force of
$M \ddot {\vec x} = - M {\vec \nabla} \Phi_{\rm ext} - \alpha Q
\vec\nabla
\varphi_{\rm ext}$ on an object of mass $M$ and charge $Q$.
It is worth emphasizing that the Vainshtein mechanism does not
suppress the scalar charge $Q$ at all -- indeed, the fact that
$Q = M$ enforces the equivalence principle, i.e. making the motion
of an object independent of its mass.
What the Vainshtein mechanism suppresses is the external
scalar field $\varphi_{\rm ext}$ that an object senses when it is
close to the source of that field.

The one exception to $Q = M$ is compact objects, 
such as black holes. These are objects whose mass
receives a significant contribution from the gravitational binding energy,
a contribution that is not included in the scalar charge $Q$.
Compact objects such as neutron stars thus have $Q/M < 1$,
with black holes as the extreme limit where $Q/M = 0$.
This is consistent with the notion of black holes having no
hair, more specifically no galileon hair which we prove in a separate
paper \cite{nohair} (see also \cite{kaloper,babichev}).
\footnote{
If one thinks of black holes as vacuum solutions, having
no scalar charge is certainly {\it a} solution, but the point
of no-hair theorem is to show that it is in fact the only
solution, and thus the collapse of an actual star would presumably lead to
such a configuration.
We should stress that our proof in \cite{nohair} concerns
only spherically symmetric black holes. We take it as suggestive
that rotating black holes likely share the same no-galileon-hair
property, but it remains to be proven. In any case,
the argument for a suppressed $Q/M$ for compact objects is
quite robust.
}
Thus, under some external $\Phi_{\rm ext}$ and
$\varphi_{\rm ext}$ fields, a normal non-compact star (as well as dark
matter particles) would fall
according to $\ddot {\vec x} = - \vec \nabla \Phi_{\rm ext}
- \alpha \vec\nabla \varphi_{\rm ext}$, while a black hole
would fall according to $\ddot {\vec x} = - \vec \nabla \Phi_{\rm
  ext}$, insensitive to the scalar force.
\footnote{The fact that compact objects have a suppressed scalar charge
is not new \cite{nordvedt,will}.
What is new with the advent of recent modified gravity models is the
screening mechanism, which revives scalar-tensor theories that are otherwise
already ruled out by solar system tests.
}

The challenge is to find situations where
the scalar $\varphi_{\rm ext}$ is not already suppressed by the
Vainshtein mechanism. Since both black holes and stars typically
reside within galaxies whose Vainshtein radii greatly exceed their sizes, it
would seem hopeless to observe the purported difference in the
rate of fall between a normal star and a black hole.

The galileon symmetry helps save the day.
The symmetry 
tells us that given any solution to the scalar Eq. (\ref{eom}), 
one can always add a linear gradient, i.e. $\varphi_{\rm ext}$ with a
constant $\vec \nabla \varphi_{\rm ext}$, and obtain another solution.
\footnote{This argument relies heavily on the galileon symmetry.
It is an interesting open question whether the offset effect between
stars and black holes would be observable in a theory like $P(X)$,
which exhibits Vainshtein screening without the galileon
symmetry.
}
For any given object, whether such a linear gradient is present
and how large it is, depends on the boundary conditions.
For a galaxy, the boundary conditions are supplied by the surrounding
large scale structure. Interestingly, as is recently demonstrated in a series of
numerical simulations \cite{koyama,fabian,mark,roman}, the galileon scalar indeed
obeys linear dynamics on sufficiently large scales ($\gsim 10$ Mpc), meaning
the large scale structure generates a galileon field that is
unsuppressed by the Vainshtein mechanism
\footnote{In other words,
while the Vainshtein mechanism does operate on small scales, the
Vainshtein zones of nonlinear objects do not percolate the
universe. Note the Vainshtein radius,
estimated from an isolated spherically symmetric object, can be a
misleading concept when applied to more complex situations.
The tensor structure of Eq. (\ref{eom}) is such that
screening works very differently in non-spherically symmetric
situations.
}.
The large-scale-structure-generated scalar field 
has a long wavelength, and can be approximated as a linear gradient
on the scale of a galaxy. 
This linear gradient penetrates the Vainshtein zone of the galaxy,
and can act unsuppressed on the galaxy and its constituents.
In other words, the galaxy falls according to this unsuppressed
scalar $\varphi_{\rm ext}$ induced by large scale structure.
So do its constituent dark matter, non-compact stars, {\it but not
its compact objects.} The most readily observable
compact object is the central massive black hole if there is nuclear activity.
The central black hole, lacking a scalar charge, does not
respond to the scalar force, while the stars (and dark matter particles)
do. The net effect is that the black hole will lag behind the stars in
their overall large-scale-structure-induced motion.
In other words, the black hole will be offset from the
center of the galaxy, or more precisely, from
the minimum of the galactic gravitational potential.
The non-zero offset 
means there's an extra (purely gravitational, not scalar) tug on the
black hole from the central region of the galaxy. This suffices
to compensate for the lack of a scalar force on the black hole,
and keep the black hole and stars in equilibrium, moving in tandem.
One can estimate the size of the offset $r$ by 
equating the extra scalar acceleration sensed by the stars
$\alpha |\vec \nabla \varphi_{\rm ext}|$ with the extra gravitational tug
on the black hole $G M_{\rm gal} (< r)/r^2$, where $M_{\rm gal} (<r)$
is the mass enclosed within radius $r$ of the galaxy.
We find a displacement
\begin{eqnarray}
\label{offset}
r = 0.1 {\, \rm kpc} \left({2 \alpha^2 \over 1}\right)
\left({|\vec\nabla \Phi_{\rm ext}| \over 20 {\rm (km/s)^2/kpc}}\right)
\left({0.01 {\rm M_\odot pc^{-3}} \over \rho_0}\right) .
\end{eqnarray}
Here, we estimate $\varphi_{\rm ext}$ by $2\alpha \Phi_{\rm ext}$,
since the linear scalar $\varphi_{\rm ext}$ satisfies 
the same Poisson equation for the gravitational potential
$\Phi_{\rm ext}$, but with the source term scaled up by $2\alpha$
(see Eq. [\ref{eom}]). The typical gravitational acceleration
$|\vec \nabla \Phi_{\rm ext}|$ can be estimated by the typical
peculiar motion multiplied by Hubble: $300 {\,\rm km/s} \times
70 {\,\rm km/s}/{\rm Mpc} \sim 20 ({\,\rm km/s})^2 / {\rm kpc}$.
A more careful calculation of the rms $|\vec \nabla \Phi_{\rm ext}|$
using the observed matter power spectrum gives a number
fairly close to this
(e.g. \cite{HG}). It should be kept in mind however
that $|\vec \nabla \Phi_{\rm ext}|$ is a stochastic quantity, 
and its value depends on environment.
The central density of $\rho_0 \sim 0.01
{\rm M_\odot /pc^3}$ is appropriate for dwarf or low surface
brightness galaxies, where the effect is the largest.
We provide explicit scaling with the relevant parameters
in Eq. (\ref{offset}) so that one can easily extrapolate
to other values.

In modeling the central region of a galaxy, several issues
should be kept in mind.
First of all, the above estimate assumes the density profile
is approximately flat close to the center. For fixed density at the core radius, a steep profile would
imply a small offset. Fortunately, low central density galaxies
where the offset is the largest also have fairly flat profiles
e.g. \cite{kratsov}. 
Second, the relevant central density is the one averaged outside
the black hole's sphere of influence. Materials within
the sphere of influence would simply move with the black hole.
It is the materials outside that are important for determining
the offset.  Thus, Bahcall-Wolf type cusps are not relevant
for our considerations \cite{bahcallwolf}.

Galaxies for which both the stars and
the central massive black hole are readily observable
are those with a low level nuclear activity, namely
Seyferts. Our estimate in Eq. (\ref{offset})
suggests that the offset would be observable
preferentially in small galaxies. 
Depending on the distance (up to say tens of Mpc), 
galaxies with a central density in the range 
$\sim 0.003 - 0.03 {\,\rm M_\odot /pc^3}$, corresponding
roughly to rotational velocities around $\sim 30 - 120$ km/s
\cite{kratsov}, should have a measurable offset.
Until recently, such small Seyferts have not been well explored
observationally.
A classic case of a dwarf galaxy containing
an active nucleus is NGC 4395 \cite{filippenko}.
A good size sample ($\sim 30$) of small Seyferts was recently
reported by \cite{moran}. \footnote{We thank Jules Halpern for pointing out
these cases to us.}

An interesting question is how fast the black hole
moves relative to the stars, on its way to the
equilibrium (offset) position. Using the same
set of parameters displayed in Eq. (\ref{offset}),
we find a fairly small velocity of $\sim 2$ km/s.
At such a velocity, dynamical friction is unimportant.
The time it takes for the black hole to traverse
the requisite distance is about $10^8$ years.
This is fairly close to some estimates of the nuclear
activity time-scale \cite{HH}, though the conditions for
nuclear activity are rather uncertain.

M87 is an interesting case, where its massive black hole
is known to be offset from the (bulge) stellar center
by a projected distance of $\sim 7$ pc \cite{M87}. 
Its central density is about 
$\rho_0 \sim 20 {\,\rm M_\odot/pc^3}$.
For our effect to reproduce such an offset thus
requires $\alpha \sim 8$, assuming the external
gravitational acceleration on M87 is typical.
This is a scalar-matter coupling that is quite a bit
stronger than gravitational.
The more plausible explanations for the observed
offset are: acceleration by an asymmetric jet, and
gravitational wave recoil from a merger \cite{M87}.
Two other explanations considered are:
the active black hole being a member of a binary, and
Brownian motion. The former can be constrained by
the lack of a jet precession, and can also be tested by
monitoring the system over time. 
The latter is a small effect, giving an offset $< 0.1$ pc
in the case of M87 \cite{M87}.

These alternative mechanisms raise a practical question: 
in the event one observes a black hole offset in a lower central
density galaxy, consistent with $\alpha \sim 1$,
how does one disentangle the modified gravity effect
from other astrophysical effects? 
One can exploit
a distinguishing feature of the modified gravity effect,
that is, the offset should be correlated with 
the gravitational
pull of the surrounding large scale structure, in both its
direction and strength. 
For instance, galaxies are expected to stream out of voids, their
resident black holes should lag behind the stars in that streaming
motion. Voids are especially interesting places to look because
the scalar field is expected to be unscreened there \footnote{A rough
argument goes as follows. 
Eq. (\ref{eom}) can be rewritten schematically in the form:
$H_0^{-2} \partial^2 \varphi + (H_0^{-2} \partial^2 \varphi)^2 \sim 
\alpha \rho/\bar\rho$. Thus, voids where $\rho/\bar\rho < 1/\alpha$
are natural
places where one can consistently ignore the nonlinear term
compared to the linear term on the left.
However, 
we expect the linear/unsuppressed scalar to be present
even away from voids, i.e. large scale structure generates a scalar
field with long wavelengths that can penetrate Vainshtein zones around
nonlinear objects.
}. However, we expect large scale
structure to source a linear (i.e. unsuppressed) scalar even away from voids.
It would be very useful to map out the precise gravitational and
scalar fields in our neighborhood, using fairly detailed knowledge
of the mass distribution of the local universe \cite{constrained}.
The two large-scale-structure-generated
fields $\Phi_{\rm ext}$ and $\varphi_{\rm ext}$ are 
expected to roughly align, but an accurate map
would aid in isolating the modified gravity effect from
other astrophysical effects.
Note also that typical astrophysical effects 
produce a {\it velocity} offset ($\sim 10 - 1000$ km/s) 
that is quite a bit larger than
what modified gravity predicts.\footnote{
An exception is Brownian motion which produces small velocity
perturbations to the massive black hole: $\sim v_{\rm star} (M_{\rm
  star}/M_{\rm b.h.})^{1/2}$. Using the relation
mass of black hole $M_{\rm b.h.} \sim 10^{8.12} {\,\rm M_\odot}
(v_{\rm star}/200
{\,\rm km/s})^{4.24}$ \cite{magorrian},
the smallest galaxies ($v_{\rm star} \sim 30$ km/s) 
have the largest velocity perturbation of
$\sim 0.2$ km/s. For such galaxies, the {\it spatial} offset
is $\sim r_{\rm core} (M_{\rm star}/M_{\rm b.h.})^{1/2} \sim 0.03$ kpc
\cite{M87}, thus comparable to the modified gravity effect.
Hence, it is important to use the correlation with large scale
structure
as a discriminant.
}

It is worth emphasizing that this offset effect -- namely
the lagging of compact objects behind stars in the
overall streaming motion of the host galaxy --
is not confined to the central massive black hole.
Any compact objects, regardless of its mass, 
will display this effect, though the effect
is larger for a black hole than a neutron star i.e.
the offset is expected to scale as $2GM/R$, where
$R$ is the radius of the object.
The advantage of the central massive black hole is that it is
readily observable even in a distant galaxy, provided it is active.
For other less massive compact objects, the best bet is to
look within our own galaxy. One possible signature is to
see if compact objects are systematically offset from 
the galactic plane (defined by the stars), in the opposite
direction of our galaxy's streaming motion.
Using numbers from Eq. (\ref{offset}), and adopting
the solar neighborhood value of $\rho_0 \sim 0.18 {\,\rm M_\odot}/{\,\rm
  pc^3}$, we estimate the offset to be about $2$ pc for black holes.
It would be useful to compute this offset more carefully
by calculating the precise large-scale-structure-generated
scalar force on the Milky Way.

A natural question is: should we expect the same offset effect 
for other screening mechanisms? 
Theories that screen by means of 
scalar self-interactions of the potential type, 
such as the chameleon \cite{chameleon} or the symmetron \cite{symmetron},
operate very differently from those that screen by
derivative interactions i.e. Vainshtein.
Non-Vainshtein theories do not have a conserved scalar charge $Q$.
Indeed, the scalar force on/from the Sun is screened in
these theories by virtue of the Sun's suppressed $Q/M$.
In chameleon and symmetron theories, objects with a 
gravitational potential similar to the Sun's ($\sim 10^{-6}$), or deeper,
have $Q/M \sim 0$. Thus, main sequence stars and black holes
fall at the same rate, and one expects no offset between the two.
However, dark matter and stars can fall
differently if the host galaxy has a sufficiently shallow
gravitational potential \cite{us}. This can lead to an interesting warping
of the galactic disk, pointed out by Jain and VanderPlas \cite{bhuv}.
For other observational tests of chameleon/symmetron screening,
see for instance \cite{chameleon,us,Brax,symmetron,phil,eugene,niayesh}.

In summary, we propose a test of Vainshtein screening
in galileon theories by comparing the rate of fall for
compact objects versus non-relativistic stars.
A positive detection of a difference will be a great boost to some of the
recent ideas of modifying gravity to explain cosmic acceleration.
A negative detection can be used to put an upper limit on
the scalar-matter coupling $\alpha$. A bound reaching $0.1$ is
conceivable with existing data, and will suffice to rule out
most of these recent ideas.
Perhaps the most interesting observation is that relatively small scale,
local data can shed light on the dark energy problem and the nature of gravity.

{\em Acknowledgements.} We would like to thank Mike Eracleous,
Bhuvnesh Jain, Christine Simpson, Erick Weinberg, 
Matias Zaldarriaga, and especially Jules
Halpern and Jacqueline van Gorkum for useful discussions. 
Our research is supported by
the DOE under contracts DE-FG02-92-ER40699 
and DE-FG02-11ER1141743, and by NASA under contract NNX10AH14G.



\begin{thebibliography}{99}

\bibitem{Bhuvreview}
  B.~Jain, J.~Khoury,
  Annals Phys.\  {\bf 325}, 1479-1516 (2010).
  [arXiv:1004.3294 [astro-ph.CO]].

\bibitem{Kurtreview}
  K.~Hinterbichler,
  [arXiv:1105.3735 [hep-th]].

\bibitem{HKW}
L. Hui, J. Khoury, J. Wang, in preparation

\bibitem{NRT}
  A.~Nicolis, R.~Rattazzi and E.~Trincherini,
  Phys.\ Rev.\  D {\bf 79}, 064036 (2009)
  [arXiv:0811.2197 [hep-th]].

\bibitem{Vainshtein}
  A.~I.~Vainshtein,
  Phys.\ Lett.\  B {\bf 39}, 393 (1972).

\bibitem{DGP}
  G.~R.~Dvali, G.~Gabadadze and M.~Porrati,
  Phys.\ Lett.\  B {\bf 485}, 208 (2000)
  [arXiv:hep-th/0005016].

\bibitem{LPR}
  M.~A.~Luty, M.~Porrati and R.~Rattazzi,
  JHEP {\bf 0309}, 029 (2003)
  [arXiv:hep-th/0303116].

\bibitem{CGT}
  C.~de Rham, G.~Gabadadze and A.~J.~Tolley,
  Phys.\ Rev.\ Lett.\  {\bf 106}, 231101 (2011)
  [arXiv:1011.1232 [hep-th]].

\bibitem{us}
  L.~Hui, A.~Nicolis and C.~Stubbs,
  Phys.\ Rev.\  D {\bf 80} 104002 (2009)
  [arXiv:0905.2966 [astro-ph.CO]].

\bibitem{us2}
 L.~Hui and A.~Nicolis,
  Phys.\ Rev.\ Lett.\ \ {\bf 105}, 231101  (2010)
  [arXiv:1009.2520 [hep-th]].

\bibitem{cristian}
 C.~Armendariz-Picon and R.~Penco,
  arXiv:1108.6028 [hep-th].

\bibitem{nohair}
L. Hui and A. Nicolis, 
companion paper.

\bibitem{kaloper}
 N.~Kaloper, A.~Padilla and N.~Tanahashi,
  JHEP\ {\bf 1110}, 148  (2011)
  [arXiv:1106.4827 [hep-th]].

\bibitem{babichev}
E. Babichev and G. Zahariade,
in preparation.
 
\bibitem{nordvedt}
 K.~Nordtvedt,
  Phys.\ Rev.\ \ {\bf 169}, 1014  (1968).

\bibitem{will}
 C.~M.~Will,
  Living Rev.\ Rel.\ \ {\bf 9}, 3  (2005)
  [gr-qc/0510072].

\bibitem{koyama} 
  A.~Cardoso, K.~Koyama, S.~S.~Seahra and F.~P.~Silva,
  Phys.\ Rev.\ D\ {\bf 77}, 083512  (2008)
  [arXiv:0711.2563 [astro-ph]].

\bibitem{fabian}
 F.~Schmidt
  Phys.\ Rev.\ D\ {\bf 80}, 043001  (2009)
  [arXiv:0905.0858 [astro-ph.CO]].

\bibitem{mark}
 J.~Khoury and M.~Wyman,
  Phys.\ Rev.\ D\ {\bf 80}, 064023  (2009)
  [arXiv:0903.1292 [astro-ph.CO]].

\bibitem{roman}
 K.~C.~Chan and R.~Scoccimarro,
  Phys.\ Rev.\ D\ {\bf 80}, 104005  (2009)
  [arXiv:0906.4548 [astro-ph.CO]].

\bibitem{HG}
 L.~Hui and P.~B.~Greene,
  Phys.\ Rev.\ D\ {\bf 73}, 123526  (2006)
  [astro-ph/0512159].

\bibitem{kratsov}
 A.~V.~Kravtsov, A.~A.~Klypin, J.~S.~Bullock and J.~R.~Primack,
  Astrophys.\ J.\ \ {\bf 502}, 48  (1998)
  [astro-ph/9708176].

\bibitem{bahcallwolf}
 J.~N.~Bahcall and R.~A.~Wolf,
  Astrophys.\ J.\ \ {\bf 209}, 214  (1976).

\bibitem{filippenko}
A. V. Filippenko and W. L. W. Sargent,
Astrophysical Journal Letters {\bf 341}, L11 (1989).

\bibitem{moran} E. C. Moran,
 Bulletin of the American Astronomical Society {\bf 42} 597 (2010)
[http://adsabs.harvard.edu/abs/2010AAS...21538301M].

\bibitem{HH}
 Z.~Haiman and L.~Hui,
  Astrophys.\ J.\ \ {\bf 547}, 27  (2001)
  [astro-ph/0002190].

\bibitem{M87} 
 D.~Batcheldor, A.~Robinson, D.~J.~Axon, E.~S.~Perlman and D.~Merritt,
  Astrophys.\ J.\ \ {\bf 717}, L6  (2010)
  [arXiv:1005.2173 [astro-ph.CO]].

\bibitem{constrained}
 S.~Gottloeber, Y.~Hoffman and G.~Yepes,
  arXiv:1005.2687 [astro-ph.CO].

\bibitem{magorrian}
F. Ferrarese and D. Merritt, ApJL {\bf 539}, 9 (2000);
K. Gebhardt et al., ApJL {\bf 539}, 13 (2000).

\bibitem{chameleon}
 J.~Khoury and A.~Weltman,
  Phys.\ Rev.\ Lett.\ \ {\bf 93}, 171104  (2004)
  [astro-ph/0309300].

\bibitem{symmetron}
 K.~Hinterbichler and J.~Khoury,
  Phys.\ Rev.\ Lett.\ \ {\bf 104}, 231301  (2010)
  [arXiv:1001.4525 [hep-th]].

\bibitem{bhuv}
 B.~Jain and J.~VanderPlas,
  JCAP\ {\bf 1110}, 032  (2011)
  [arXiv:1106.0065 [astro-ph.CO]].

\bibitem{Brax} 
  P.~Brax, C.~Burrage, A.~-C.~Davis, D.~Seery and A.~Weltman,
  Phys.\ Rev.\ D {\bf 81}, 103524 (2010)
  [arXiv:0911.1267 [hep-ph]].

\bibitem{phil}
 P.~Chang and L.~Hui,
  Astrophys.\ J.\ \ {\bf 732}, 25  (2011)
  [arXiv:1011.4107 [astro-ph.CO]].

\bibitem{eugene}
 A.~-C.~Davis, E.~A.~Lim, J.~Sakstein and D.~Shaw,
  arXiv:1102.5278 [astro-ph.CO].

\bibitem{niayesh}
 R.~Pourhasan, N.~Afshordi, R.~B.~Mann and A.~C.~Davis,
  arXiv:1109.0538 [astro-ph.CO].

\end{thebibliography}
\end{document}